\documentstyle[11pt]{article}
\begin{document}

\def\be{\begin{equation}}
\def\fe{\end{equation}}
\def\eqn{\label}
\font\sm=cmr9
\tolerance=5000


\title{\bf RELATIVISTIC DYNAMICS OF VORTEX DEFECTS IN SUPERFLUIDS}

\author{ BRANDON CARTER\\
D\'epartement d'Astrophysique Relativiste et de Cosmologie,\\
Centre National de la Recherche Scientifique, \\Observatoire de
Paris, 92195 Meudon, France.}

\date{February, 1999.}
\maketitle

\noindent
{(\sm Lecture notes for Les Houches winter school
 TOPOLOGICAL DEFECTS AND NON - EQUILIBRIUM DYNAMICS OF PHASE TRANSITIONS, 
ed Y. Bunkov, H. Godfrin.)}
\bigskip



{\bf Abstract:} Superfluid condensates are known to occur in contexts
ranging from laboratory liquid helium to neutron stars, and are also
likely to occur in cosmological phenomena such as axion fields. In the
zero temperature limit, such condensates are describable at a
mesoscopic level by irrotational configurations of simple relativistic
perfect fluid models.  The general mechanical properties of such models
are presented here in an introductory review giving special attention
to the dynamics of vorticity flux 2-surfaces and the action principles
governing both individual flow trajetories and the evolution of the
system as a whole.  Macroscopic rotation of such a condensate requires
the presence of a lattice of quantised vortex defects, whose averaged
tension violates perfect fluid isotropy. It is shown that for any
equation of state (relating the mass density $\rho$ to the pressure
$P$) the mesoscopic perfect fluid model can be extended in a uniquely
simple and natural manner to a corresponding macroscopic model (in a
conformally covariant category) that represents the effects of the
vortex fibration anisotropy.  The limiting case of an individual vortex
defect is shown to be describable by a (``global'') string type model
with a variable tension ${\cal T}$ (obtained as a function of the
background fluid density) whose ``vorton'' (i.e. closed loop
equilibrium) states are derived as an exercise.

\section{Introduction.}
\label{Section 1}

These lectures offer an introduction to the dynamics of vortices in
simple superfluid models of a general category that includes the kind
appropriate in the context of laboratory condensed matter for the
representation of Helium-4 in the zero temperature limit (for which no
``normal'' entropy carrying constituent is present), as well as the
ultrarelativistic ``stiff'' kind that has been widely
used~\cite{VilenkinVacha87} for the representation of a massless axion
field in cosmology. Between these extremes, this category also includes
the kind of model suitable as a simple approximation for the description of
the neutron-pair condensation superfluid that is generally believed
(for both theoretical and observational reasons~\cite{Sauls88}) to
occur in the intermediate layers of neutron stars. (A more sophisticated 
treatment would include allowance for the presence of an interpetrating 
ionic crust lattice or of an independently superconducting proton pair 
condensate, not to mention complications such as spin.) 

The declared purpose of this school for which these lectures are
intended is to study analogies between low temperature laboratory
phenomena and high energy cosmological phenomena. However it is
desirable, when possible, to reinforce mere analogy by interpolation. I
would therefore like to emphasize the particular interest for this
purpose of the intermediate regime of neutron star interiors, about
which a great deal of admittedly indirect information is available from
analysis of pulsar frequency variations.  In contrast with most other
areas of astrophysics, but in common with typical laboratory
applications of condensed matter physics, the temperatures in typical
neutron star interiors can usually be considered to be very low
compared with the relevant energy scales. On the other hand, in
contrast with typical laboratory applications of condensed matter
physics, but in common with high energy cosmological scenarios, the
treatment of neutron star interiors requires allowance for significant
relativistic effects.

While evidently indispensible for an accurate treatment of neutron 
star matter, a relativistic treatment is commonly -- but wrongly --
considered to be an unnecessary complication for low temperature
laboratory applications. For a reasonably accurate treatment of
laboratory liquid Helium a relativistic treatment is indeed
unnecessary, but what is wrong is to suppose that it is more
complicated.  On the contrary, as I hope these lectures will make
clear, the relativistic treatment, in so far as it is available, is
mathematically simpler.  (The essential reason for this is that the
Lorentz group is, in the technical sense ``semi-simple'' whereas the
Galilei group is not.)

The work, in collaboration with David Langlois~\cite{CarterLanglois95},
on which these lectures are based, was originally inspired by the
observation by Rick Davis and Paul Shellard~\cite{DavisShellard89} (in
precisely the spirit that is the official raison d'\^etre for the
present school) of a strong analogy between the behaviour of vortices
in the ``stiff'' (ultrarelativistic) massless axion model at one extreme, 
and in the incompressible superfluid model that is commonly used for the
description of zero temperature Helium-4 at the opposite extreme. 

What is shown here is the way to carry out the analogous, but not
quite so trivially simple, treatment that is needed for the
intermediate category of generically compressible superfluid models.
At a mesoscopic level these models will be of irrotational perfect
fluid type, and will be characterised by a subluminal speed, $c_{_{\rm
I}}$ say, of ordinary ``first'' sound, that will be determined -- by an
equation of state specifying the pressure $P$ as a function of the mass
density $\rho$ -- according to the familiar formula 
\be c_{_{\rm I}}=\big(dP/d\rho\big)^{1/2}\, .\eqn{0a}\fe 
Rotation at a macroscopic level will entail the presence of an
Abrikosov type lattice of quantised vortex defects whose averaged
tension produces a deviation from perfect fluid isotropy.  Analysis of
individual vortex tubes~\cite{CarterLanglois95b} indicates that, in the
limit for which their relative ``drift'' or ``flight'' velocity is
highly subsonic, their averaged effect can be described by the
inclusion of an extra isotropy violating term of a uniquely simple and
natural kind~\cite{CarterLanglois95} in the action for a corresponding
macroscopic model.  Despite their violation of isotropy, such models
(for different equations of state) will be  shown to belong to a
category that is conserved by conformal transformations. More
particularly, it will be shown that the specific model corresponding to
the high pressure limit of a gas of negligibly interacting particles as
characterised by $c_{_{\rm I}}^{\,2} $ $=c^2/3$ (where $c$ is the speed
of light) has the same kind of conformal invariance property as the
well known example of Maxwell's equations.

The scope of the present review does not extend to the  recently
perfected relativistic analogue~\cite{CarterKhalatnikov92,CarterLanglois95a} 
of Landau's original non-dissipative two constituent superfluid model
(which has recently been shown to be expressible in a very elegant Galilean
covariant form~\cite{CarterKhalatnikov94}) for the purpose of allowing
for the dynamical effect of the ``normal'' entropy flux current that
would be present at a non zero temperature). Such effects are not  the
ones that are most important in the context of neutron stars, the main
application for which the work described here is intended, where the
temperature will typically be so low (compared with the relevant M.e.v.
range energy scales) that thermal corrections can for most purposes be
neglected.  For application to neutron stars, a more important kind of
generalisation is to a relativistic models capable of describing
neutron superfluid penetration~\cite{Carter89a} of the solid material
forming the crust, and of describing ~\cite{CarterLanglois98}  the
protonic superconductivity that is expected~\cite{Sauls88} within the
neutron superfluid at a deeper level below the crust.

\section{Canonical treatment of relativistic flow trajectories.}
\label{Section 2}

Before proceeding it is desirable to recall some essentials the
relativistic kinematics and dynamics, particularly  in view of the
regretable tradition in non-relativistic fluid theory -- and most
notably in non-relativistic superfluid theory -- of obscuring the
essential distinction between velocity (which formally belongs in a
tangent bundle) and momentum (which formally belongs in a cotangent
bundle) despite the fact that the distinction is generally respected in
other branches of non-relativistic condensed matter theory, such as
solid state physics, where the possibility of non-alignment between the
3-velocity $v^a$, and the effective 3-momentum $p_a$ of an electron
travelling in a metallic lattice is well known.

In a non relativistic treatment it is only in strictly Cartesian
(rather than e.g. cylindrical or comoving) that the distinction between
contravariant entities such as the velocity $v^a$ and covariant
entities such as the momentum $p_a$ can be ignored.  In a relativistic
treatment, even using coordinates $x^\mu\leftrightarrow \{t,x^a\}$ of
Minkowski type, with a flat spacetime metric $g_{\mu\nu}$ whose
components are of the fixed standard form diag$\{-c^2,\,1,\,1,\,1\}$,
(where $c$ is the speed of light) the necessity of distinguishing
between raised and lowered indices is inescapable.  Thus for a
trajectory parametrised by proper time $\tau$, the correponding unit
tangent vector
\be u^\mu={d x^\mu\over d\tau}\eqn{1}\fe
is automatically, by construction a contravariant vector: its space
components, $u^a=\gamma v^a$ with $\gamma=(1-v^2/c^2)^{-1/2}$ will be
unaffected by the index lowering operation $u^\mu\mapsto
u_\mu=g_{\mu\nu}u^\nu$, but its time component $u^{_0}=dt/d\tau=\gamma$ will
differ in sign from the corresponding component $u_{0}=-\gamma c^2$ of the
associated covector $u_\mu$. On the other hand the 3-momentum $p_a$ and energy
$E$ determine a 4-momentum covector $\pi_\nu$ that is intrinsically covariant,
with components $\pi_a=p_a$, $\pi_{_0}=-E$ . The covariant nature of the
momentum can be seen from the way it is introduced by the defining equation,
\be \pi_\nu={\partial L\over \partial u^\nu}\, ,\eqn{2a}\fe in terms of the
relevant position and velocity dependent Lagrangian function $L$, from which
the corresponding equation of motion is obtained in the well known form
\be {d\pi_\nu\over d\tau}={\partial L\over\partial x^\nu}\, .\eqn{2b}\fe

In the case of a free particle trajectory, and more generally for fluid flow
trajectories in all the  simple ``barotropic'' perfect fluid models with which
the present lectures will be concerned, the Lagrangian function will have the
familiar standard form \be L={_1\over^2}\mu g_{\mu\nu}u^\mu
u^\nu-{_1\over^2}\mu c^2 \, ,\eqn{3}\fe in which (unlike what is needed for
more complicated chemically inhomogeneous models\cite{Carter79,Carter89a}) it
is the same scalar spacetime field $\mu$ that plays the role of mass in the
first term and that provides the potential energy contribution in the second
term. The momentum will thus be given by the simple proportionality relation
\be \pi_\nu=\mu u_\nu\, ,\eqn{3a}\fe
so that one obtains the expressions $E=\gamma\mu c^2$, $p_a=\mu\gamma v_a$, in 
which the field $\mu$ is interpretable as the relevant effective mass.

In the case of a free particle model, the effective mass $\mu$ will of
course just be a constant, $\mu=m$. This means that if, as we have been
supposing so far, the metric $g_{\mu\nu}$ is that of flat Minkowski type, the
resulting free particle trajectories will be obtainable trivially as straight
lines. However the covariant form of the equations (\ref{1}) to (\ref{3a})
means that they will still be valid for less trivial cases for which, instead
of being flat, the metric $g_{\mu\nu}$ is postulated to have a variable form
in order to represent the effect of a gravitational field, such as that of a
Kerr black hole (for which, as I showed in detail in a much earlier Les
Houches school~\cite{Carter73}, the resulting non trivial geodesic equations
still turn out to be exactly integrable).

In the case of the simple perfect fluid models with which we shall be concerned
here, the effective mass field $\mu$ will be generically non-uniform.
In these models the equation of state giving the pressure $P$ as a function
of the mass density $\rho$ can most conveniently be specified by first
giving $\rho$ in terms of the corresponding conserved number density $n$
by an expression that will be decomposible in the form
\be \rho =mn+{\epsilon\over c^2}\, ,\eqn{6}\fe
in which $m$ is a fixed ``rest mass'' characterising the kind of particle
(e.g. a Cooper type neutron pair) under consideration, while $\epsilon$
represents an extra compression energy contribution. The pressure will
then be obtainable using the well known formula
\be P=(n\mu-\rho) c^2\, ,\eqn{7}\fe
in which the effective dynamical mass $\mu$ (or equivalently the
``specific enthalpy'' $\mu c^2$) is given by
\be \mu={d\rho\over dn}=m+{1\over c^2}{d\epsilon\over dn}\, .\eqn{8}\fe
It is this that is to be taken as the effective mass function appearing
in the specification (\ref{3}) of the relevant Lagrangian.

When one is dealing not just with a single particle trajectory but a
spacefilling fluid flow, it is possible and for many purposes desirable to
convert the Lagrangian dynamical equation (\ref{2b}) from particle evolution
equation to equivalent field evolution equations~\cite{Carter79,Carter89a}. 
Since the momentum covector
$\pi_\nu$ will be obtained as a field over spacetime, it will have a well
defined gradient tensor $\nabla_{\!\rho} \pi_\nu$ that can be used to rewrite
the left hand side of (\ref{2b}) in the form 
$d\pi_\nu/d\tau=u^\rho\nabla_{\!\rho}\pi_\nu$. Since the value of the Lagrangian
will also be obtained as a scalar spacetime field $L$, it will also have a
well defined gradient which will evidently be given by an
expression of the form
$\nabla_{\!\nu} L={\partial L/\partial x^\nu}+\big({\partial L/\partial
\pi_\rho}\big)\nabla_{\!\nu}\pi_\rho\, . $
We can thereby rewrite the Lagrangian dynamical equation (\ref{2b}) as a field
equation of the form
\be u^\rho\nabla_{\!\rho}\pi_\nu+\pi_\rho\nabla_{\!\nu}u^\rho=\nabla_{\!\nu}L
\, .\eqn{12}\fe

An alternative approach is of course to start from the corresponding
Hamiltonian function, as obtained in terms of the position and momentum
variables (so that formally it should be considered as a function on the
spacetime cotangent bundle) via the Legendre transformation
\be H=\pi_\nu u^\nu-L\, .\eqn{13}\fe
In this approach the velocity vector (\ref{1}) and the dynamical
equation (\ref{2b}) are recovered using the familiar formulae
\be {dx^\mu\over d\tau}={\partial H\over \partial \pi_\nu}\, , \hskip 1 cm
{d\pi_\nu\over d\tau}=-{\partial H\over\partial x^\nu}\, .\eqn{14}\fe
The consideration that we are concerned not just with a single trajectory but
with a spacefilling fluid means that, as in the case of the preceding
Lagrangian equations, so in a similar way this familiar Hamiltonian dynamical
equation can also be converted to a field equation which takes the form 
\be  2 u^\rho\nabla_{\![\rho}\pi_{\nu]}=-\nabla_{\!\nu} H\, ,\eqn{15}\fe 
with the usual convention that square brackets are used to indicate index
antisymmetrisation. On contraction with $u^\nu$, the left hand side
will evidently go out, leaving the condition
\be u^\nu\nabla_{\!\nu}H=0 \, ,\eqn{16}\fe
expressing the conservation of the value of the Hamiltonian along the
flow lines.

The actual form of the Hamiltonian function that is obtained from the
particularly simple kind of Lagrangian function (\ref{3}) with which
we are concerned will evidently be given by
\be H={1\over 2\mu} g^{\nu\rho}\pi_\nu \pi_\rho+{\mu c^2\over 2}
\, .\eqn{17}\fe
In order to ensure the proper time normalisation for the parameter
$\tau$ the equations of motion (in whichever of the four equivalent
forms (\ref{2b}), (\ref{12}), (\ref{14}), (\ref{15}) may be preferred)
are to be solved subject to the constraint that -- 
in order for $u^\mu$ to be correctly normalised -- the numerical value of
the Hamiltonian should vanish,
\be H=0\hskip 1 cm \Rightarrow \hskip 1 cm u^\mu u_\mu=-c^2\, ,\eqn{18}\fe
initially , and hence also by (\ref{16}) at all other times. In the
more general systems that are needed for some purposes the Hamiltonian
may be constrained in a non uniform manner~\cite{Carter79,Carter89a} 
so that the term on the right of (\ref{15}) will be non zero, but in
the simpler systems that suffice for our present purpose the restraint
(\ref{18}) ensures that this final term will drop out, leaving a
Hamiltonian equation of the very elegant and convenient form
\be u^\mu \nabla_{[\mu}\pi_{\nu]}=0\, .\eqn{20}\fe

\section{Vorticity conservation and flux 2-surfaces.}
\label{section 3}

The preceding form (\ref{20}) of the dynamical equations is particularly 
handy for the analysis of symmetries and the derivation of conservation 
laws according to which physically interesting quantities are preserved
by various kinds of continuous displacement~\cite{Carter79,Carter89a}. 
The variation induced by an infinitesimal displacement generated
by an arbitrary vector field $k^\mu$ say, will be given
by the corresponding Lie derivative operator, whose effect on a scalar
field, $\mu$ say, will evidently be given simply by
\be\vec k{\cal L}\mu=k^\nu\nabla_{\!\nu}\mu\, ,\eqn{21}\fe
while its effect on the metric will be given by the well known though
not quite so immediately obvious formula
\be \vec  k{\cal L} g_{\mu\nu}=2\nabla_{\!(\mu}k_{\nu)}\, ,\eqn{22}\fe
using round brackets to denote index antisymmetrisation. If $k^\mu$ is a 
Killing vector field, i.e. if the displacement generated by $k^\mu$ is
a symmetry of the spacetime metric, then  the right hand side of (\ref{22}) 
will vanish. In curved space such symmetries are rare, but in ordinary flat 
space timethere is of course a ten parameter family of such Killing vector 
fields generating the Poincar\'e group, whose algebra has as its basis the four
independent generators of uniform spacetime translations and the six
independent generators of the Lorentz group.  The effect of the Lie
differentiation operation on another vector field, $u^\rho$, say will be given
simply by their mutual commutator bracket,
\be \vec k{\cal L}u^\rho =\big[\vec k,\vec u\big]{^\rho}
=-\vec u{\cal L}k^\rho\, ,\hskip 1 cm
\big[\vec {k},\vec {u}\big]{^\rho}=k^\nu\nabla_{\!\nu}u^\rho
-k^\nu\nabla_{\!\nu}k^\rho\, .\eqn{23}\fe
Such Lie differentiation is an example of an operation for which the distinction
between covectors and ordinary contravariant vectors is important: except when
performed with respect to a Killing vector, Lie differentiation does not
commute with index raising and lowering. The rule that applies to the covector
$u_\mu=g_{\mu\nu}u^\nu$ can evidently be obtained by combining (\ref{22}) and
(\ref{23}), so that for $\pi_\nu$ one obtains a formula that can be
conveniently expressed in terms of exterior (antisymmetrised) derivatives in
the form
\be\vec k{\cal L}\pi_\nu=2k^\rho\nabla_{[\rho}\pi_{\nu]}
+\nabla_{\!\nu}(\pi_\rho k^\rho)\, .\eqn{24}\fe
Although it does not commute with index raising and lowering, Lie
differentiation does commute with exterior differentiation. Thus for the
relativistic vorticity tensor, which is defined to be the antisymmetrised
derivative of the momentum covector, i.e.
\be w_{\mu\nu}=2\nabla_{[\!\mu}\pi_{\!\nu]}\, ,\eqn{25}\fe
so that its own exterior derivative will automatically vanish, i.e.
\be \nabla_{\![\mu}w_{\nu\rho]}=0\, ,\eqn{26}\fe
it follows that its Lie derivative will be obtainable just by taking 
the exterior derivative of (\ref{24}) which gives
\be \vec k{\cal L}w_{\mu\nu}=-2\nabla_{[\mu}(w_{\nu]\rho} k^\rho)
\, .\eqn{27}\fe

The preceding general formula (\ref{24}) can immediately be used to rewrite
the Lagrangian dynamical equation (\ref{12}) in the expressive 
form~\cite{Carter79,Carter89a}
\be\vec u{\cal L}\pi_\nu=\nabla_{\!\nu}L\, ,\eqn{28}\fe
from which by taking the exterior derivative, one can immediately derive the
dynamical vorticity conservation law that is the key to superfluidity theory
(and much else) in the form
\be \vec u{\cal L}w_{\mu\nu}=0 \, .\eqn{29}\fe
The interpretation of this crucially important result is that for any
flow governed by the Lagrangian equations (\ref{2b}) or their Hamiltonian 
equivalent (\ref{14})  (which have so far simply been postulated
ex cathededra, but whose validity for any ``barotropic'' perfect
fluid be made clear in the following section) the vorticity
field (\ref{25}) will simply be convected onto itself by the
flow field $u^\mu$, with the implication that if it vanishes initially
$w_{\mu\nu}$  will remain zero throughout the flow, which in this case will
be describable as ``irrotational''.

The foregoing results can be considerably strengthenned in cases such as those
of the simple ``barotropic'' perfect fluids considered here for which the
proper time normalisation is ensured by the Hamiltonian constraint (\ref{18}).
This has the effect of reducing the dynamical equations to the particularly
simple form (\ref{20}), whose interpretation is that the flow vector $u^\mu$
must be an zero eigenvalue eigenvector of the vorticity tensor $w_{\mu\nu}$.
The posession of a zero eigenvalue requires that $w_{\mu\nu}$ should
satisfy the degeneracy condition
\be  w_{\mu[\nu}w_{\rho\sigma]}=0\, ,\eqn{31}\fe
which excludes the possibility of it having matrix rank 4, with the implication
that unless it actually vanishes it must have rank 2 (since an antisymmetric
tensor can never have odd integer rank). This means that the flow vector
$u^\mu$ is just a particular case within a whole 2-dimensional tangent
subspace of eigenvectors $e^\mu$ satisfying
\be e^\mu w_{\mu\nu}=0\, .\eqn{32}\fe
This subspace will be spanned by a unit worldsheet element tangent
bivector ${\cal E}^{\mu\mu}$ of the kind whose use was developped
 by Stachel~\cite{Stachel80}, and that is definable,
wherever the vorticity magnitude 
\be w=\big({_1\over^2} w_{\mu\nu}w^{\mu\nu}\big)^{1/2} \eqn{32a} \fe
does not vanish, as being proportional to the dual vorticity tensor
$W^{\mu\nu}$, i.e.
\be {\cal E}^{\mu\nu}= {1\over w}W^{\mu\nu}\, ,\hskip 1 cm
W^{\mu\nu}={1\over 2}\varepsilon^{\mu\nu\rho\sigma}\,
w_{\rho\sigma} ,\eqn{33a}\fe
(note that sign convention used for the worldsheet orientation here is the
opposite of what was used in the preceding article~\cite{CarterLanglois95})
where $\varepsilon^{\mu\nu\rho\sigma}$ is the totally antisymmetric tensor
normalised by the convention that its non zero components are equal to 1 or -1
(depending on whether the index ordering is an even or odd permutation) with
respect to locally Minkowskian coordinates with $g_{\mu\nu}\leftrightarrow$
${\rm diag}\{-c^2,\,0,\, 0,\, 0\}$, so that with respect to an
arbitrary coordinate system the non zero component values of
$\varepsilon^{\mu\nu\rho\sigma}$ will  be given by $\pm c\Vert g\Vert^{-1/2}$.

It can be seen that this worldsheet tangent bi-vector will satisfy
\be {\cal E}^{\mu\nu}{\cal E}_{\mu\nu}=-2c^2\, ,
\hskip 1 cm  {\cal E}^{\mu\nu}w_{\nu\rho}=0\, .\eqn{35}\fe
It follows from this last equation that the contraction of any
covector with ${\cal E}^{\mu\nu}$ will provide a solution of
the vortex worldsheet tangentiality condition (\ref{32}). 
A noteworthy example is the helicity $h^\mu$ as defined
\cite{Carter79,CarterKhalatnikov92} by
\be h^\mu=w{\cal E}^{\mu\nu}\pi_\nu  \, ,\eqn{36}\fe
which can be seen from (\ref{26}) and (\ref{31}) to satisfy the 
helicity current conservation law~\cite{Carter79,Carter89a}
\be \nabla_{\!\nu}h^\nu=0\, .\eqn{36a}\fe

Again using the degeneracy condition (\ref{31}) and the condition that the
vorticity also satisfies the Poincar\'e closure condition (\ref{26}), it can be
shown \cite{Carter89a} that the tangent elements 
characterised by ${\cal E}^{\mu\nu}$ and generated by solutions of
(\ref{32}) (or more specifically by $u^\nu$ and $h^\nu$)  will
automatically satisfy the relevant Frobenius condition for integrability,
meaning that they will mesh together to form well behaved timelike
2-dimensional worldsheets.  This makes it possible to extend the flow
line conservation laws resulting from any continuous symmetries that
may be present.

The simplest example is the Bernouilli type theorem that applies --
even if the Hamiltonian does not satisfy the constraint (\ref{18}) --
whenever $k^\rho$ is a symmetry generator of the system, so that in
particular $\vec k{\cal L}H$ and $\vec k{\cal L}\pi_\nu$ both vanish:
it can be seen from from (\ref{15}) and (\ref{24}) that
$u^\nu\nabla_\nu(\pi_\rho k^\rho)$ will vanish, so $\pi_\rho k^\mu$
will be constant along each flow line. In the most obvious 
application, the Killing vector is just the generator of time
translations in flat space with the Minkowski coordinates, 
$k^\rho\leftrightarrow \{1,\,0,\,0,\,0\}$, so that the
Bernouilli constant will simply be identifiable with the negative of
the effective energy per particle, i.e we shall have $\pi_\rho k^\rho=
-E$ with $E=\gamma\mu c^2$ which automatically includes allowance for
both compression energy and kinetic energy contributions.

This conclusion can be greatly strengthenned when the Hamiltonian constraint
(\ref{18}) holds: one then gets
$e^\nu\nabla_\nu(\pi_\rho k^\rho)=0$ for any vector $e^\nu$ satisfying
(\ref{32}) which evidently means that $\pi_\rho k^\nu$ will not just be
constant along each flow trajectory but that it will be constant throughout
each of the 2-dimensional vorticity flux worldsheets. In the irrotational case
\be w_{\mu\nu}=0 \, ,\eqn{40}\fe
for which (\ref{32}) is satisfied trivially by any vector
$e^\nu$ at all, one can draw the even stronger conclusion that $\pi_\rho
k^\rho$ will be constant throughout the fluid:
\be \vec k{\cal L}\pi_\rho= 0\ \ \Rightarrow \ \ \nabla_{\!\nu}(\pi_\rho 
k^\rho)=0\, .\eqn{41}\fe
(A similar conclusion of global uniformity of $\pi_\rho k^\rho$ would also be
obtainable immediately from (\ref{18}) and (\ref{24}) in the alternative, more
widely familiar, case of motion that is rigid in the sense of having a flow
vector that is aligned with the Killing vector, i.e. for which $u^{[\nu}
k^{\rho]}=0$.) It is to be emphasised that the applicability of this kind of
generalised Bernouilli theorem is not limited to the case of ordinary
stationarity, for which $k^\rho$ is a time translation generator, but is just
as well applicable to cases of axisymmetry for which the Killing vector is a
rotation generator, and it has recently been found to be very
useful~\cite{Gougetal99} for a hybrid case (involving a non-rigidly rotating
binary pair of tidally deformed and thus non-axisymmetric neutron stars which,
from the point of view of a distant observer, are not stationary but
periodically evolving, but which are nevertheless stationary from the point of
view of a local observer with respect to a suitably rotating frame).

An alternative way of obtaining the local vorticity conservation theorem
(\ref{29}) is as the differential limit of the global Kelvin-Helmoltz 
theorem to the effect that, as is manifest from the form (\ref{28})
of the Lagrangian dynamical equation, the action integral
\be S=\int\pi_\nu dx^\nu \eqn{42}\fe will be preserved if taken round a closed
circuit that is convected by the flow field $u^\mu$. In the irrotational case
(\ref{40}), the Jacobi type action integral (\ref{42})
between any two fixed endpoints will be unaffected by continuous displacements
of the path between them, and hence can be used to construct a locally
well defined field $S$ that is not only such that one has 
\be \pi_\nu=\nabla_\nu S\, ,\eqn{43}\fe but that
will also be a solution of the Hamilton Jacobi equation specified by setting
the Hamiltonian function to zero with the gradient of S substituted in place
of the momentum.

The special case of a (simple, zero temperature limit) superfluid is specified
by the existence of a well defined mesoscopic phase factor ${\rm
e}^{i\varphi}$ (representing the phase factor of an underlying bosonic
condensate that might consist of Helium-4 atoms or Cooper type neutron pairs)
in which the phase angle $\varphi$ is given according to the usual
correspondence principle by
\be \varphi=S/\hbar\, .\eqn{43a}\fe
In a multiconnected configuration of a classical irrotational fluid the Jacobi
action field  $S$ obtained from (\ref{43}) might have an arbitrary
periodicity, but in a superfluid there will be a U(1) quantisation
requirement that the periodicity of the phase angle $\varphi$ should be a
multiple of $2\pi$, and thus that the periodicity of the Jacobi action $S$
should be a multiple of $2\pi \hbar$. The simplest configuration for any such
superfluid is a uniform stationary state in a flat Minkowski background, for
which the phase will have the standard plane wave form 
\be S/\hbar =k_a x^a-\omega t\, ,\eqn{44}\fe
from which one obtains the correspondence $\pi_\nu\leftrightarrow
\{-\hbar \omega, \hbar k_a\}$, which means that the effective
energy per particle will be given by $E=\gamma\mu c^2=\hbar\omega$
and that the 3-momentum will be given by $p_a=\mu\gamma v_a=\hbar k_a$.

It is to be remarked that for ordinary timelike superfluid particle
trajectories the corresponding phase speed $\omega/k$ of the wave
characterised by (\ref{44}) will always be superluminal, a fact which
people working with liquid  Helium-4 in the laboratory can blithely
ignore, since what matters for most practical purposes is not the phase 
speed but the group velocity of perturbation wave packets. Our present
discussion will be limited to the strict zero temperature limit for
which no such packets are excited, but it is easy to extend the
relativistic analysis to low but non-zero temperatures for which the
relevant excitations are phonons~\cite{CarterLanglois95a}. Although
their  phase speed and group velocity are the same, both being given by
the formula (\ref{0a}) for the ordinary (``first'') soundspeed which
will of course be subluminal, phonons do nevertheless have a tachyonic
aspect of their own: their 4-momentum covector is always spacelike, in
contrast with that of an ordinary fluid or superfluid particle which is
timelike. This means that whereas the effective energy $E$ of an
ordinary fluid or superfluid particle is always positive, the effective
energy $E$ of a phonon may be positive or negative, depending on
whether the frame of reference with respect to which it is measured is
moving subsonically or supersonically. The well known implication is
that if the superfluid is in contact with a supersonically moving
boundary there will inevitably be an instability giving rise to
dissipative phonon creation.

\section{Conventional formulation of perfect fluid and simple superfluid theory.}
\label{Section 4}

Although sufficient for the derivation of many important properties 
of the flow, the dynamical equations on which the preceding
sections have been based contain only part of the information  needed
for a complete determination of the perfect fluid evolution. The most usual
way of presenting the complete set of equations of motion of a simple perfect
fluid of the barotropic type we are considering -- meaning one whose intrinsic
local physical state is characterised just by a single independent scalar
field -- is in the form of a conservation law of
the standard form
\be \nabla_{\!\nu}T^{\mu\nu}=0\, ,\eqn{45}\fe
for a stress momentum energy density tensor that is specified as a
function of the timelike unit flow tangent vector $u^\nu$ and a single
independent scalar field variable, such as the conserved particle
number density $n$, on which the other relevant quantities, such as the
effective mass $\mu$ given by (\ref{8}), will be functionally
dependent.

For a system of this simple kind, the 4 independent components of the
energy momentum conservation law (\ref{45}) provide all that is needed
to determine the evolution of the 4 independent components that
characterise the local state of the system, which can be taken to be
the scalar $n$ and the 3 space components $u^a$ of $u^\mu$ (since
the remaining component $u^{_0}$ is not dynamically independent but
determined by the unit normalisation condition (\ref{18}) as an
algebraic function of the 3 other components).

In any perfect fluid model the mass density $\rho$ and pressure $P$ are
physically characterised by their role in the specification of the
stress momentum energy density tensor, for which the standard
expression is \be T^{\mu\nu}=(\rho+{P\over c^2})u^\mu u^\nu +P
g^{\mu\nu}\, .\eqn{46}\fe
In the simple barotropic case, the relation (\ref{7}) between the
dependences of $P$ and $\rho$ on $n$  can be seen to be necessitated by
the requirement that the dynamical system (\ref{45}) should ensure
conservation of the number current
\be n^\mu=n u^\mu\, .\eqn{47}\fe
It is easy to check, using (\ref{7}) that contraction of (\ref{47})
with $u_\mu$ does indeed lead to the required result, namely
\be \nabla_{\!\mu} n^\mu=0\, .\eqn{48}\fe
One can also verify the not quite so well known result~\cite{Lichnerowicz67} 
that the remaining independent equations
(\ref{45}) can be reorganised in the canonical uniformly Hamiltonian 
form (\ref{20}) on which the work~\cite{Carter79,Carter89a} of the 
preceding subsection was based, and which is expressible succinctly as
\be n^\mu w_{\mu\nu}=0\, .\eqn{49}\fe
Thus in addition to the formula (\ref{49}) that has been used so far,
the only additional information needed for the complete specification
of the dynamics of a barotropic fluid system is the obvious particle
conservation law (\ref{48}).

Given a dynamical system, one of the first things any physicist is inclined 
to ask is whether it is derivable from a Lagrangian type variation principle. 
We have already seen in the previous sections that (\ref{49}) by itself is
obtainable from Lagrangian equations of motion for the individual
trajectories, which are of course obtainable from a one dimensional action
integral of the form $\int L\, d\tau $ with $L$ as given by (\ref{3}). The
question to be adressed now is how to obtain the complete set of dynamical
equations (\ref{45}), including (\ref{48}) as well as (\ref{49}), from an
action integral over the 4-dimensional background manifold 
${\cal S}^{^{(4)}}$ of the form
\be {\cal I}=\int {\cal L}\, d{\cal S}^{^{(4)}}\, , \hskip 1 cm 
d{\cal S}^{^{(4)}}= {\Vert g\Vert^{1/2}\over c} d^4 x\, , \eqn{50}\fe
for some suitable scalar Lagrangian functional ${\cal L}$.

Several radically different procedures are available for doing this.
Although ultimately equivalent ``on shell'', they involve variation
over ``off shell'' bundles that differ not just in structure but even
in dimension.  The oldest and most economical from a dimensional point
of view is the worldline variation procedure developed by
Taub\cite{Taub54}, followed Clebsch type variation procedure developed
by Schutz\cite{Schutz70}, but for our present purpose it will be more
convenient to employ the more recently developed  Kalb-Ramond type
method\cite{Carter94} that has been specifically designed for dealing
with problems of macroscopic superfluidity.

The problem is greatly simplified if, to start off with, one restricts
oneself to the purely irrotational case (\ref{40}),
which is all that is needed for the description of zero temperature
superfluidity at a mesoscopic level. For this case
independent variable can be taken to be just the Jacobi action $S$,
or equivalently in a superfluid context, the phase $\varphi$ as given by
(\ref{43a}), and the action is simply taken to be the pressure P
expressed as a function of the effective mass $\mu$, with the latter
constructed as proportional to the amplitude of the 4-momentum, 
according to the prescription
\be \mu^2 c^2=-\pi_\nu \pi^\nu \eqn{52}\fe
with the 4-momentum itself given by the relation (\ref{43}) that
applies in the irrotational case, i.e.
\be \pi_\nu=\hbar\nabla_\nu\varphi\, .\eqn{53}\fe
Thus setting
\be {\cal L}=P\, ,\eqn{54}\fe
and using the standard pressure variation formula
$ \delta P= c^2 n \delta \mu$
one sees that the required variation of the Lagrangian will be
given by
\be \delta{\cal L}=-n^\nu\delta\pi_\nu=-\hbar n^\nu\nabla_{\!\nu}(\delta
\varphi)\, .\eqn{54b}\fe
Demanding that the action integral (\ref{50}) be invariant with
respect to infinitesimal variations of $\varphi$ then evidently leads
to the required conservation law (\ref{48}). 

\section{Introduction of the dilatonic amplitude field $\Phi$.}
\label{Section 5}

For an equation of state such that $P\propto\mu^2$, the Lagrangian
(\ref{54}) will, as it stands, have the quadratic field gradient
dependence that is typical of simple physical field theories. This
occurs in the special limit case, relevant to the massless axion field
in cosmology~\cite{VilenkinVacha87,DavisShellard89}, of the ``stiff''
model, characterised by (\ref{66}) as discussed below, for which the
``first'' sound speed (\ref{0a}) is equal to that of light.  Except in
this idealised limit case, a feature of the preceding Lagrangian
function (\ref{54}) that is widely considered to be undesirable is its
generically non quadratic dependence on $\nabla\varphi$, an apparent
drawback that is not uncommonly dealt with by recourse to
approximation~\cite{Davis90}.

What this secion will show however is that that, not just for $P\propto\mu^2$
but for a quite general equation of state, the Lagrangian (\ref{54}) 
can be reformulated in the much more desirably fashionable form 
\be {\cal L}=-{\hbar^2\over 2}\Phi^2(\nabla_{\!\nu}\varphi)
\nabla^\nu\varphi-V\{\Phi\} \eqn{56}\fe
-- in which the potential energy density term $V$ is some suitably chosen
algebraic function of the amplitude $\Phi$ -- by a transformation of variables
that is absolutely exact~\cite{Carter94}, without any need for recourse to
approximation provided one adopts the correct definition for the auxiliary
field variable $\Phi$.

What would involve an approximation would be to to implement a further step
whereby $\Phi$ is ``promoted''
from the status of an auxiliary variable to that of an extra dynamical
variable by adding in a supplementary kinetic term,
\be \Delta{\cal L}_{\{a\}}=-a^2\hbar^2(\nabla^{\,\nu}\Phi)\nabla_{\,\nu}\Phi
\, ,\eqn{57}\fe
for some constant value of $a$. Various kinds of gradient term, of
which this is the most obvious, might indeed be added for the purpose
of improving the physical precision of the model in regimes of rapid
amplitude variation, where deviations from the strictly isotropic
perfect fluid form (\ref{46}) might be expected to become significant.
However for a strongly interacting liquid like Helium-4 (as opposed to
a weakly interacting gas) I know of no reliable procedure for the
quantitative evaluation of such a term on the basis of an underlying
many-particle quantum theory.  The most commonly used~\cite{Davis90}
ansatz, namely to take $a=1$, is not automatically guaranteed to
provide an improvement but might even bring about a deterioration of
physical precision in some circumstances. The standard choice $a=1$
does not have theoretical or empirical foundations but is merely based
on the purely mathematical consideration that it provides an
adjusted Lagrangian
${\cal L}_{\{1\}}={\cal L} +\Delta_{\{1\}} {\cal L}$
of the ``relativistic Ginzburg Landau'' form
\be {\cal L}_{\{1\}}=-{\hbar^2\over 2}(\nabla_{\!\nu}\Psi)
\nabla^\nu \bar\Psi-V\{|\Psi|\}\, ,\eqn{58}\fe
in which the phase variable $\varphi$ and the amplitude variable
$\Phi$ have been combined to form a complex variable
\be \Psi={\rm e}^{i\varphi}\, ,\hskip 1 cm  \bar \Psi={\rm e}^{-i\varphi}
\, ,\hskip 1 cm \Phi=|\Psi| \, .\eqn{59}\fe
This feature is useful for some purposes -- notably for providing a
smoothed out treatment~\cite{BenYa91,BenYa92} of the vortex defects
that arise where the phase $\varphi$ becomes indeterminant -- but not
so convenient for deriving exact results such as the uniformity of the
Bernouilli constant $\pi_\nu k^\nu$ that was shown above to be valid
for a model of the original kind whenever $k^\rho$ is a symmetry
generator.  

As far as physical accuracy (as opposed to mathematical convenience) is
concerned it will usually not matter very much whether one uses the
exact perfect fluid model given by (\ref{56}) or the associated Ginzburg  
Landau model given by (\ref{58}) since their difference, as given by
the extra kinetic term (\ref{57}) can be expected to be very small
except in the immediate neighbourhood of a vortex core where neither
model can be expected to be physically accurate. What does matter for
physical accuracy is the choice of the functional form for the
amplitude $\Phi$.  

When it comes to the point, without even attempting to provide any
serious microscopic derivation, many introductory presentations start
by assuming the validitity of the Landau Ginzburg framework, and then
resort to crude guesswork for the specification of $\Phi$, typically
taking $\Phi\propto \sqrt{n}$, or even more commonly 
\be \Phi \propto \sqrt{\rho} \, . \eqn{61}\fe 
Such a choice just happens to provide a quantitatively acceptable
result in the non-relativistic limit, but the flimsiness of its
theoretical basis is exposed by the fact that for a generic
relativistic model it gives the wrong answer in the regime of slow
amplitude variation where the (correctly calibrated) Landau Ginzburg
model (\ref{58})  should agree with the relevant perfect fluid model
(\ref{56}). 
The correct choice -- the only way to get exact
agreement between (\ref{54}) and (\ref{56}) -- is to take
\be \Phi={n\over\sqrt{\rho+{P/ c^2}}}=\Big({n\over\mu}\Big)^{1/2}\,
 ,\eqn{62}\fe
the corresponding characterisation for the potential energy density
function $V$ being~\cite{Carter94} that it should be given by
\be V={\rho c^2-P\over 2}\, .\eqn{63}\fe

When characterised in this manner, -- with the correct identification
(\ref{62}) rather than (\ref{61})  --
 the perfect fluid model given by the Lagrangian (\ref{56}) provides a
uniquely canonical model for the representation at zero temperature of
a simple superfluid such as Helium-4, at least in the regime where the
rate of variation of the amplitude $\Phi$ is not too rapid compared
with that of the phase. It is to be emphasised that for this regime the
justification for this perfect fluid model is actually on a sounder
footing than that of the corresponding Ginzburg Landau type model
(\ref{58}), since appart from the postulate of the isotropic perfect
fluid form (\ref{46}) for the stress momentum enery density tensor,
whose applicability in the slow amplitude variation regime is hard to
doubt, and the invocation of the generally valid conservation laws
(\ref{45}) and (\ref{47}), the only assumption on which this fluid
model is based, and for which it relies on an underlying microscopic
quantum analysis, is that of the rather well established existence of
the mesoscopic phase $\varphi$.

Having evaluated $V$ as a function of $\Phi$ one can recover the
effective mass $\mu$, number density $n$, mass density $\rho$
and pressure $P$ of the fluid using the formulae
\be \mu^2={1\over c^2\Phi}{dV\over d\Phi}\, ,\hskip 1 cm n=\Phi^2 \mu\, ,
\eqn{63a}\fe
and 
\be \rho={_1\over^2}\,\Phi^2\mu^2+{V\over c^2}\, ,\hskip 1 cm
P={_1\over^2}\,\Phi^2\mu^2 c^2 -V\, ,\eqn{63b}\fe
which are derivable from (\ref{7}) and (\ref{8}).

Since the preceding relations entail the variation rule
\be \delta V=\mu^2\,c^2\,\Phi\, \delta\Phi\, ,\eqn{64}\fe
it can be seen that the full variation of the reformulated perfect
fluid Lagrangian (\ref{56}) will be given by
\be \delta{\cal L}=-\big(\pi_\nu \pi^\nu+\mu^2 c^2 \big)\Phi\delta\Phi
-\hbar\Phi^2\pi^\nu\nabla_{\!\nu} (\delta\varphi)\, ,\eqn{65}\fe
with the momentum covector $\pi_\nu$ as specified before by (\ref{53}).
Thus in this new formulation, instead of imposing (\ref{52}) as a defining
relation, we obtain it as a field equation from the requirement that the
action integral should be invariant with respect to local variations of the
auxiliary field $\Phi$, while as before the analogous requirement for
variations of the phase variable $\varphi$ gives back the particle
conservation law (\ref{47}).

\subsection{Special equations of state}
\label{Section 5a}

It is to be remarked that whereas for a generic compressible fluid equation of
state giving $\rho$ as a function of $n$, and hence giving $P$ as a function
of $\mu$, the formula (\ref{64}) will provide a corresponding function $V$
that provides a reformulated action function of the form (\ref{56}) from which
the required perfect fluid dynamical equations are obtainable by treating
$\Phi$ and $\varphi$ as independent. There is however an exceptional case that
works somewhat differently, namely that of the ``stiff'' Zel'dovich model
-- pertaining to the massless axion field  in
cosmology~\cite{VilenkinVacha87,DavisShellard89} -- which is characterised by 
\be P=\rho c^2\, ,\eqn{55}\fe
corresponding by (\ref{0a}) to sound propagation at the speed of
light.  This case is obtained from a primary equation of state of the
form $\rho\propto n^2$ or equivalently (as observed at the beginning of
Section \ref{Section 5}) $P\propto\mu^2$, which gives $V=0$ and $\mu\propto
n$. This last relation means that the amplitude $\Phi$ will
simply be a constant, and so will not act as an auxiliary variable in
the usual way. In this case what one has to do to obtain the relevant
``stiff'' dynamical equations is simply to treat $\varphi$ as the only
independent variable in the Lagrangian (\ref{56}).

There is another important special limit for which it is not the new
formulation but the original variational formulation (\ref{54}) in terms of
the pressure function that fails to work, namely the extreme case -- which
often useful as an approximation --  of a ``dust type'' model characterised by
$\rho\propto n$ for, which there is no pressure, i.e. $P=0$. This low pressure
limit model (the only model for which the naive identification (\ref{61}) 
is exactly valid) is immediately obtainable within the framework of the new 
formulation (\ref{56}) simply by taking $V\propto \Phi^2$. 

The simplest non trivial potential energy function that can be used in
the new formulation (\ref{56}) is provided by the ``radiation gas''
model (\ref{55}) characterised by 
\be  c_{_{\rm I}}^{\, 2}= {c^2\over 3}\hskip 1 cm \Rightarrow \hskip 1 cm
V\propto\Phi^4\, ,\eqn{66}\fe 
which is obtainable from a primary equation of state of the form
$\rho\propto n^{4/3}$. This is the model that is appropriate for
representing the cosmological black body radiation (which is to a good
approximation irrotational, though of course it is not a superfluid).
It also applies to a degenerate Fermi gas of massless (or due
to high compression effectively massless) non interacting particles,
and for that reason was used as a first crude approximation in some of
the pioneering studies of neutrons stars, whose intermediate layers are
indeed believed to be superfluid. As will be explained below, this
particular model is characterised by conformal invariance of the same kind
as is familiar in the well known case of Maxwellian electromagnetism.

For more general purposes, as an approximation that can be usually be expected
to be reasonably accurate within a limited density range, one can of course
use the obvious generalisation of (\ref{66}) to the standard form
\be V={m^2 c^2\over2}\Phi^2+a^2\Phi^4\, ,\eqn{67}\fe
for suitably adjusted constants $m$ and $a$. For such a model
the fluid mass density and pressure will be given according to
(\ref{63b}) by
\be \rho=m^2\Phi^2+{3 a^2\over c^2}\Phi^4\, ,\hskip 1 cm
P={a^2\Phi^4}\, ,\eqn{68}\fe
from which it can be seen that the sound speed (the quantity one would
probably want to use in practice to fix the appropriate value of $a$)
will be given according to (\ref{0a}) by
\be c_{_{\rm I}}^{\,2}=\Big({3\over c^2}
+{m^2\over 2 a^2\Phi^2}\Big)^{-1} \eqn{69}\fe

\section{Introduction of the Kalb Ramond gauge field}
\label{Section 6}

The treatment given in the preceding section is ideal for the description of
a superfluid at a mesoscopic (i.e. intervortex) scale, but for the treatment
of an ordinary perfect fluid with rotation, or for the treatment of a
superfluid on a macroscopic scale (allowing for the averaged effect of a large
number of vortices) more general models are required.

A first step towards the kind of generalisation that is needed is to 
formulate the current in terms of an antisymmetric Kalb Ramond type tensor
field $B_{\mu\nu}=-B_{\nu\mu}$ whose exterior derivative
\be N_{\mu\nu\rho}=3\nabla_{\![\mu}B_{\nu\rho]}\, ,\eqn{70}\fe
whose (Hodge type) dual
\be n^\mu={1\over 3!}\varepsilon^{\mu\nu\rho\sigma}N_{\nu\rho\sigma}
\eqn{71}\fe
is to be identified with the particle number current (\ref{47}), which will
evidently be invariant under the effect of Kalb  Ramond gauge transformations
$B_{\mu\nu}\mapsto B_{\mu\nu}+2\nabla_{\![\mu} \chi_{\nu]}$. It can then be
seen that the consequent closure condition,
\be \nabla_{[\mu}N_{\nu\rho\sigma]}=0\, ,\eqn{72}\fe
is equivalent to the usual form (\ref{48}) of the particle conservation law.

The idea now is to perform a Legendre type transformation ${\cal L}\mapsto
\Lambda$ whereby the independent scalar field $\varphi$ of the preceding
formulation based on ${\cal L}$ is replaced by the antisymmetric gauge tensor
$B_{\mu\nu}$ in a new formulation of the same model in terms of a different
dually related Lagrangian function $\Lambda$ which takes the form
\be \Lambda=-{c^2\over 12\Phi^2}\, N^{\mu\nu\rho} N_{\mu\nu\rho}
-V\{\Phi\}\, .\eqn{73}\fe
In the preceding formulation based on ${\cal L}$ as given by (\ref{56}) the
irrotationality property was kinematically imposed in advance while the
particle current conservation was obtained from the variational principle as 
a dynamical equation. However in the reformulated version based on the dual
Lagrangian (\ref{73}) it is the particle current conservation law that is
obtained in advance as we have seen via the kinematic identity (\ref{72}),
while on the other hand the irrotationality condition (\ref{40}), i.e.
\be \nabla_{\![\mu}\pi_{\nu]}=0\, , \eqn{74}\fe
is obtained directly from (\ref{73}) in the equivalent dual form
\be \nabla_{\!\nu}\big(\Phi^{-2} N^{\nu\rho\sigma}\big)=0\, .\eqn{75}\fe
from the requirement of invariance with respect to independent
variations of the gauge 2-form $B_{\mu\nu}$.

The purpose of replacing the simple scalar field $\varphi$ by the 
tensorial field $B_{\mu\nu}$ is to enable the extension of the model to 
the general perfect fluid case, in which the particle conservation law 
(\ref{72}) is retained, but the irrotationality condition (\ref{74}) is
abandoned. The way to do this~\cite{Carter94} is to introduce a complete
Lagrangian of the form
\be {\cal L}=\Lambda-{_1\over^4}\varepsilon^{\mu\nu\rho\sigma}
B_{\mu\nu}w_{\mu\nu}\, ,\eqn{76}\fe
where, in analogy with what has already been done for the current 3-form, the
vorticity 2-form is constructed from independent gauge fields in such a way
that its conservation property, (\ref{26}) is automatically ensured in advance
as a kinematic identity. For this purpose the requisite independent gauge
fields can be taken to be a pair of independent scalars, $\chi^\pm$ say, in
terms of which an identically conserved vorticity flux will be given by
\be w_{\nu\sigma}= 2\big(\nabla_{\![\nu}\chi^+\big)\nabla_{\!\sigma]}
\chi^-\, .\eqn{77}\fe

When the Lagrangian (\ref{76}) is substituted in the action integral 
(\ref{50}), the requirement of invariance with respect to local variations of 
the (non-physical) dynamical gauge fields $B_{\mu\nu}$ and $\chi^\pm$, and of 
the (physical) auxiliary amplitude $\Phi$, can be seen to lead back to our 
original dynamical momentum transport equation (\ref{49}). Thus (since the 
particle conservation law (\ref{48}) has been imposed kinematically in advance) 
it provides the complete system of generic perfect fluid equations of motion 
as given by the standard stress momentum energy conservation law (\ref{45}).

\section{Macroscopic allowance for vortex quantisation.}
\label{Section 7}

All we have done so far is to reformulate ordinary barotropic perfect fluid
theory in such a way that the vorticity $w_{\mu\nu}$ comes in as an
independent dynamical variable determined by the pair of scalar gauge fields
$\chi^\pm$. The model describing simple zero temperature superfluidity is
obtained within this framework simply by taking the scalars $\chi^\pm$ to be
constants (e.g. zero) so as to get $w_{\mu\nu}=0$.

For a macroscopic treatment of a bulk superfluid fibred by a dense congruence
of discrete vortex defects, the need for an extra term in the action to allow
for the extra tension and energy in the vortices was recognised long
ago~\cite{BekarevichKhalat61} in the laboratory context of Helium-4, and has
more recently been taken into account in the Newtonian mechanical analysis of
neutron star matter~\cite{Mendell91}. In an analogous manner, following
earlier work ~\cite{LebedevKhalat82} on the corresponding relativistic
formulation needed for a more accurate treatment of neutron star matter, the
more complete treatment~\cite{CarterLanglois95} to be described here takes
account of the averaged effect of quantised vortices aligned in an Abrikosov
type lattice by summing over contributions of individual vortex cells as
estimated~\cite{CarterLanglois95b} using the usual (very good) approximation
in which the hexagonal cells are treated as if they were cylindrically
symmetric. What this analysis suggests is that the averaged effect of such
vortices can be represented rather well -- provided any relative flow is
highly subsonic -- by a remarkably simple and mathematically elegant
modification of the generic potential function introduced by (\ref{63}): all
that seems to be necessary is to make an adjustment of the very simple form
 \be V\{\Phi\}\mapsto V\{\Phi\}+\Phi^2\Upsilon\{w\}\, ,\eqn{78}\fe
for some function $\Upsilon$ depending just on the scalar magnitude $w$ of the
vorticity as given by (\ref{32a}). Furthermore, as was observed in analogous
investigations in a non relativistic framework~\cite{BekarevichKhalat61}, the
dependence on the vorticity magnitude is approximately linear, having the form
\be \Phi^2\Upsilon=\lambda w\, ,\eqn{78b}\fe
in which
\be \lambda ={\cal K}\Phi^2\, ,\hskip 1 cm \Upsilon={\cal K} w\, ,\eqn{79}\fe
with a coefficient ${\cal K}$ that can be taken to be a constant of the order
of the Planck value, $ {\cal K} \approx \hbar$.

A remarkable consequence of the ansatz (\ref{78}) in conjunction with the
linearity postulate (\ref{79}) is that, as will be described below, the
extended model retains the noteworthy, though little known, conformal
convariance property of the simple perfect fluid model.

A more specific quantitative estimate of the value of the coefficient in
(\ref{79}) is given by
\be {\cal K}={\hbar\,\hat l\over 4}\, ,\hskip 1 cm
\hat l={\rm ln}\,\big\{ {\Delta^2\over\delta^2}\big\}
\, ,\eqn{81}\fe
where $\delta$ is an inner cut off length representing the microscopic vortex
core radius, and $\Delta$ is a long range cut of length typically representing
the mean intervortex separation distance.

For a very precise treatment one would of course need to allow for a weak
logarithmic dependence of ${\cal K}$ on $w$ since $\Delta^2$ will be inversely
proportional to $w$, but so long as $\Delta$ is very large compared with
$\delta$, as will be the case in typical macroscopic applications, the effect
of such a refinement will in practice be negligible. To obtain very high
precision when the Mach (flow to sound speed) ratio is non-negligible
~\cite{CarterLangloisPriou97} one might also have to allow for some sort of
tensorial, not just scalar dependence on the vorticity tensor.

Assuming that a sufficiently accurate treatment is obtainable without the need
to take account of tensorial vorticity dependence,  it follows that the
complete macroscopic superfluid Lagrangian will be 
given~\cite{CarterLanglois95} by
\be {\cal L}=-{c^2\over 12\Phi^2}N^{\mu\nu\rho}N_{\mu\nu\rho} -V\{\Phi\}
+{_1\over ^2}\big({\lambda\over c^2}
{\cal E}_{\mu\nu}-B_{\mu\nu}\big)W^{\mu\nu}\, ,\eqn{83}\fe
where $W^{\mu\nu}$ is the dual vorticity vector as defined by (\ref{33a}).

In order to obtain the equations of motion for this system, one needs to
evaluate the corresponding variation which will be expressible by 
\be \delta L= \pi_\nu \delta n^\nu -{_1\over^2}w_{\mu\nu}
\delta b^{\mu\nu}-{_1\over^2}\big(b^{\mu\nu}+\lambda^{\mu\nu}\big)
\delta w_{\mu\nu}+\big({c^2 n^2\over\Phi^3}-{dV\over d\Phi}-{2\lambda w
\over \Phi}\Big)\delta\Phi \eqn{84a}\fe
using the notation
\be b^{\mu\nu}={_1\over^2}\varepsilon^{\mu\nu\rho\sigma}B_{\rho\sigma}\, ,
\hskip 1 cm \lambda^{\mu\nu}={\lambda\over w}
w^{\mu\nu}\, .\eqn{84b}\fe
Requiring invariance of the corresponding integral with respect to $\delta
B_{\mu\nu}$, or equivalently to $\delta b^{\mu\nu}$, just leads back again to
the usual relation (\ref{25}) specifying the vorticity 2-form $w_{\mu\nu}$ as
the exterior derivative of the momentum 1-form $\pi_\nu$, so there will be a
conserved helicity vector given by exactly the same formula (\ref{36}) as
before. The equation of state relation (\ref{63a}) is however modified by the
addition of an extra term proportional to the vorticity: requiring invariance
with respect to $\delta \Phi$ gives
\be \Phi^2\mu^2 c^2={n^2 c^2\over\Phi^2}=\Phi{dV\over d\Phi}+2\lambda w
\, .\eqn{82a}\fe
The final requirement is invariance with respect to variations of the gauge
scalars determining the vorticity field according to (\ref{77}), which 
is equivalent to the gauge independent requirement of invariance with 
respect to Lie transportation, $\delta w_{\mu\nu}=\vec k{\cal L}w_{\mu\nu}$, 
as given by (\ref{27}) for an arbitrary displacement vector field
$k^\mu$. The resulting dynamical equation is expressible in the form
\be ^\sharp\!n^\mu w_{\mu\nu}=0\, ,\eqn{82b}\fe
which differs from its analogue in the perfect fluid limit only by the
replacement of the conserved particle current $n^\mu$ by an  ``augmented''
current   $^\sharp\!n^\mu$ that is also automatically conserved
\be \nabla_{\!\mu}\, ^\sharp\! n^\mu=0\,  ,\eqn{83c}\fe
whose specification is given by
\be  ^\sharp\!n^\mu= n^\mu+\nabla_{\!\nu}\lambda^{\mu\nu}\, ,\eqn{83d}\fe
which is equivalent, by (\ref{84b}) , to the condition that it be obtainable
from a corresponding ``augmented'' Kalb-Ramond gauge 2-form given by
\be ^\sharp\! B_{\mu\nu}= B_{\mu\nu} -{\lambda\over c^2}{\cal E}_{\mu\nu}
\, .\eqn{82e}\fe
The geometrical interpretation of the modified dynamical equation (\ref{82b})
is facilitated by the observation that it is the same as the single vortex
equation of motion (\ref{113}) that will be derived below.

\section{The conformal covariance property}
\label{Section 8}

It is well known that -- when expressed in the usual way in terms of an
electromagnetic gauge potential $A_\nu$ and charge current $j^\mu=e n^\mu$,
where $e$ is a charge coupling constant and $n^\mu$ is an automatically
conserved current vector such as is obtainable from a corresponding closed
3-form $N_{\mu\nu\rho}$ according to the Hodge type duality formula
(\ref{71}) -- the equations of ordinary Maxwellian electromagnetism are
preserved by any conformal transformation of the form
\be g_{\mu\nu}\mapsto \tilde g_{\mu\nu}={\rm e}^{2\phi} g_{\mu\nu}
\, ,\eqn{C1}\fe
for an arbitrary scalar field $\phi$, on the understanding that the
transformation affects neither the gauge 1-form, which obeys $A_\nu\mapsto
A_\nu$, nor the closed current 3-form, which obeys $N_{\mu\nu\rho}\mapsto
N_{\mu\nu\rho}$. This last condition means that, since (\ref{C1}) implies
\be {\varepsilon}^{\mu\nu\rho\sigma}\mapsto \tilde
 {\varepsilon}^{\mu\nu\rho\sigma}= 
{\rm e}^{-4\phi}{\varepsilon}^{\mu\nu\rho\sigma}\, ,\eqn{C2}\fe
the current vector itself will undergo a conformal transformation
of the form
\be n^\mu\mapsto \tilde n^\mu= {\rm e}^{-4\phi} n^\mu\, .\eqn{C3}\fe
In the framework of the Kalb-Ramond representation (\ref{70}) this is evidently
equivalent to the requirement of preservation of the gauge 2-form, 
\be B_{\mu\nu}\mapsto \tilde B_{\mu\nu}=B_{\mu\nu}\, .\eqn{C4}\fe

What is not so well known (since the utility of the auxiliary field $\Phi$, as
correctly defined by (\ref{62}), is not yet widely appreciated) is that the
ordinary barotropic perfect fluid equations (\ref{48}) and (\ref{49}) are also
preserved by such a conformal transformation, subject to the understanding
that the auxiliary amplitude field undergoes a corresponding conformal
transformation of the form
\be \Phi\mapsto\tilde\Phi={\rm e}^{-\phi}\Phi\, ,\eqn{C5}\fe
which is what is required to ensure the preservation, 
\be \pi_\nu\mapsto \tilde\pi_\nu=\pi_\nu\, ,\eqn{C6}\fe
of the momentum 1-form $\pi_\mu$ (whose role in the perfect fluid case is
analogous to that of the gauge 1-form $A_\mu$ in the electromagnetic case).  

The preservation of the form of the dynamical equations (\ref{48}) and
(\ref{49}) is in general not quite sufficient for preservation of the 
complete system, because it is also necessary to satisfy the algebraical
equation of state relation (\ref{63a}) specified by the function
$V\{\Phi\}$ which governs the relation between the amplitude $\Phi$ and
the 4-momentum magnitude $\mu=(-\pi^\nu\pi_\nu)^{1/2}$. However provided 
the conformal scalar $\phi$ is chosen to depend only the amplitude $\Phi$, 
making the later a function of new variable $\tilde\Phi$, then the system 
will be formally covariant in the sense that the new system will also behave 
as an ordinary barotropic fluid but with a modified equation of state
function $\tilde V\{\tilde\Phi\}$ in place of the orginal potental 
$V\{\Phi\}$.

My original discussion of the perfect fluid case~\cite{Carter94} envisaged a
scenario involving gravitational coupling in the framework of a 
Brans-Dicke-Jordan type generalisation of Einstein's theory,
with the dimensionless field $\phi={\rm ln}\{\tilde\Phi/\Phi\}$ acting 
as a dilatonic coupling scalar. Conformal covariance of this scheme was
found to require that the transformation law for $V$ should be given by an 
algebraic relation of the form $\tilde V/\tilde\Phi^4=V/\Phi^4$. (For readers 
 interested in this scenario I should warn that it is necessary to 
the correct the final sentence, concerning a special case for which there is a 
transformation to a form in which ``the dilatonic field is genuinely absent'': 
the term ``genuinely'' should be replaced by ``apparently'', since although 
it disappears from the fluid sector of the Lagrangian it effectively turns up
again in the gravitational sector instead.)

In present discussion I wish to describe a kind of conformal covariance 
like that of the well known Maxwellian example, having nothing to do with any 
particular kind of gravitational coupling theory, whether it be that of 
Einstein or anyone else. In order to preserve the formal structure of the fluid 
system by itself (without involving anything to do with active gravitational
coupling) it can be seen from (\ref{63a}) that since we shall have
\be \mu\mapsto \tilde\mu={\rm e}^{-\phi}\mu\, ,\eqn{C7} \fe
it is necessary and sufficient that the effective potential should transform
in such a way as to satisfy the differential condition
\be  \tilde\Phi^{-3}{d\tilde V\over d\tilde\Phi}=\Phi^{-3}{d V\over d\Phi}
\, .\eqn{C8}\fe

This means, for example, that if the new system is to be 
characterised by a pressure free ``dust'' type equation of state of 
the form $\tilde V={_1\over^2}\tilde m^2 c^2 \tilde\Phi^2$ for some 
constant $\tilde m$ (which has the convenient feature that the flow 
trajectories will simply be geodesics with respect to the new metric 
$\tilde g_{\mu\nu}$) then the desired transformation giving $\tilde\Phi$ 
as a function of $\Phi$ will be obtainable immediately by taking
$\tilde m^2 c^2\tilde\Phi^{-2}=\Phi^{-3}dV/d\Phi$
except in the exceptional case for which the right hand side of
(\ref{C8}) is constant.

Whereas a perfect fluid solution for a generic equation of state can thus be
conformally transformed to a solution for any other generic equation of state,
there is an exceptional case for which such a procedure fails, namely the 
case with  a constant value for right hand side of (\ref{C8}), which is that
of the ``radiation gas'' model for which the dependence on $\Phi$ of the
effective potential $V$ is of the homogeneously quartic form (\ref{66}). It is
apparent that the particular form 
\be P={\rho c^2\over 3}\hskip 1 cm \Rightarrow\ \hskip 1 cm
\mu \propto \Phi \fe 
of the equation of state relation for this particular model will
automatically be preserved by (\ref{C5}) and (\ref{C7}), not just when
the conformal scalar $\phi$ is chosen as a function of $\Phi$ but for
any field $\phi$ whatsoever.

In summary, just as solutions of Maxwell's equation are well known to be 
mapped conformally onto other solutions of Maxwell's equations, solutions of 
the equations of the ``radiation gas'' model  are analogously mapped 
conformally onto other solutions of the ``radiation gas'' model, while 
generically solutions of the barotropic perfect fluid equations for any 
equation of state can be conformally mapped onto solutions for another 
different equation of state. In view of the intrinsic isotropy of a perfect 
fluid these conformal properties are not very startling. However we conclude 
this subsection by the rather more surprising observation that these properties 
will still hold, subject to the same condition (\ref{C8}), when the intrinsic
isotropy is violated by the inclusion of the term (\ref{78b}) that allows for
the energy and tension in quantised vortices. Since we shall have
\be {\cal E}_{\mu\nu}\mapsto \tilde {\cal E}_{\mu\nu}= 
{\rm e}^{2\phi}{\cal E}_{\mu\nu}     \, ,\eqn{C9a}\fe
and
\be w\mapsto\tilde w={\rm e}^{-2\phi} w\, ,\eqn{C9}\fe 
it can be seen that in order to have invariance of $^\sharp\!B_{\mu\nu}$ as 
given by (\ref{82e}) and of the extra term $\lambda w$ in the action integral
(remembering that the measure $d{\cal S}^{(4)}$ will transform proportionally 
to ${\rm e}^{4\phi}$) we must have \be\lambda\mapsto\tilde\lambda=
{\rm e}^{-2\phi}\lambda\, .\eqn{C10}\fe 
Rather remarkably, this condition just happens to be satisfied by the 
formula (\ref{79}), but it  would fail for a more complicated
vorticity dependent term such as might be needed for an accurate treatment of
relative flow at a speed comparable with that of sound.

\section{The thin vortex string limit}
\label{Section 9}

Let us now consider the limiting case for which instead of being
4-dimensionally extended, the vorticity distribution is concentrated in the
neighbourhood of some particular vorticity flux 2-surface ${\cal S}^{(2)}$
say, which might conveniently be characterised by zero values for the scalar
gauge fields $\chi^\pm$. Such a 2-surface will be describable in terms of
internal coordinates $\sigma^{_0}$ and $\sigma^{_1}$, which might conveniently
be specified by the values on the worldsheet of the spacetime coordinates $t$ 
and $x^{_1}$.

However the internal coordinates may be chosen, the worldsheet embedding
$\{\sigma^{_0},\, \sigma^{_1}\}\mapsto x^\mu=\bar x^\mu \{\sigma\}$ will
induce (in the technical sense ``pull back'') a two surface metric with
components $\gamma_{ab}$ on the worldsheet that will be given by
\be \gamma_{ab}=g_{\mu\nu}\bar x^\mu_{\, ,a}\bar x^\nu_{\, ,b}\, ,
\eqn{85}\fe
using a comma to denote partial differentiation with respect to the internal
coordinates $\sigma^a$ ($a=0,\, 1$), and this induced metric in turn will
specify the worldsheet measure
\be d{\cal S}^{^{(2)}}={\Vert \gamma\Vert^{1/2}\over c}\, 
d\sigma^{_0}\,d\sigma^{_1}\, ,\eqn{85b}\fe
of the timelike 2-surface element spanned by the coordinate variations
$d\sigma^{_0}$ and $d\sigma^{_1}$.

Let us start by supposing that the vorticity distribution is confined within a
small but finite range specified by displacements $\delta
\chi^+,\,\delta\chi^-$ of the comoving scalar fields, and then let us take the
thin string limit as the size of these displacements tends to zero. In this
limit it can be seen that the dual vorticity (\ref{33a}) will take the form of
a two dimensional Dirac distribution that will be expressible in Dirac's
notation as an integral over the 2-dimenssional world sheet ${\cal S}^{(2)}$
by the formula
\be  W^{\mu\nu}={c\over \Vert g\Vert^{1/2}}\int \overline W{^{\mu\nu}}
\delta^4[x^\mu-\bar x^\mu\{\sigma\}]\, d{\cal S}^{^{(2)}}\, ,\eqn{87}\fe
in which $\overline W{^{\mu\nu}}$ is a well behaved antisymmetric tensor on
the worldsheet representing the concentrated 2 surface vorticity flux. 

In the case of the continuous vorticity distribution considered in the
preceding section, the vorticity conservation law (\ref{26}) is expressible
in its dual version in the form
\be \nabla_{\!\nu}W^{\mu\nu}=0\, .\eqn{88}\fe
When one goes over to the singular limit in which the distribution
$W^{\mu\nu}$ is concentrated in the form (\ref{87}) in the infinitesimal
neighbourhood of a single flux worldsheet, the conservaton law (\ref{88}) will
translate into a corresponding condition on the regular worldsheet supported
2-surface vorticity flux tensor $\overline W{^{\mu\nu}}$. Using the
abbreviation
\be \overline \nabla_{\!\mu}=\eta^\nu_{\ \mu}\nabla_{\! \nu}
\eqn{89}\fe
for the operator of worldsheet tangential covariant differentiation, where
$\eta^\nu_{\ \nu}$ is the ``first'' fundamental
tensor~\cite{Carter89b,Carter95} of the worldsheet (i.e. the rank-2 projection
operator whose contraction with an abritrary vector at a point on the
worldsheet projects it onto its tangential part within the worldsheet) which
will be given by
\be \eta^\nu_{\ \nu}=c^{-2}{\cal E}^{\nu\rho}{\cal E}_{\rho\mu}
\, ,\eqn{90}\fe
the condition expressing the conservation of the vorticity flux
distribution in the worldsheet will expressible simply as
\be \overline\nabla_{\!\nu}\overline W{^{\mu\nu}}=0\, .\eqn{91}\fe
Since the canonically normalised worldsheet tangential bivector
${\cal E}^{\mu\nu}$ automatically satifies a conservation
condition of the same form
\be \overline\nabla_{\!\nu}{\cal E}^{\mu\nu}=0\eqn{92}\fe
as a kinematic identity, it follows that (\ref{91}) is interpretable as
implying that the worldsheet supported surface vorticity flux tensor
$\overline W^{\mu\nu}$ must have the form
\be \overline W{^{\mu\nu}}=\kappa{\cal E}^{\mu\nu}\, ,\eqn{93}\fe
where $\kappa$ is a constant on the worldsheet, i.e.
$\overline\nabla_{\!\nu}\kappa=0$.

It can be seen that the constant $\kappa$ defined by (\ref {93})
will be interpretable as the value in the thin string limit of the
2-surface integral of the vorticity across any small spacelike section 
through the world tube. By Stoke's theorem, using the defining
relation (\ref{25}), this will be equal to the value of the Jacobi action
around the boundary circuit of the section, which can be taken to be
any closed curve encircling the vortex string. Thus for any
such surrounding circuit we shall have
\be \kappa=\oint dS=\oint\pi_\nu dx^\nu= 2\pi\hbar\nu\, ,\eqn{94}\fe
where $\nu$ is an integer representing the winding number of 
the phase $\varphi$, i.e. $\nu$ represents the number of individual
quantised vortices carrying the flux under consideration. This means 
that if we are considering a string representing just a single such 
quantised vortex, with orientation chosen so that $\nu=+1$
we shall simply have 
\be \kappa =2\pi\hbar\, .\eqn{94b}\fe

Substituting (\ref{83}) in formula (\ref{50}) it can be seen that one 
obtains the total action integral in the form
\be {\cal I}={\cal I}_{_{(4)}}+{\cal I}_{_{(2)}}\, , \eqn{95}\fe
in which the first part is just the 4-dimensional contribution from the 
irrotational superfluid outside the vortex, which by  (\ref{73}) is
\be {\cal I}_{_{(4)}}=\int {\cal L}_{_{(4)}}\, d{\cal S}^{^{(4)}}\, ,\hskip 
1 cm {\cal L}_{_{(4)}}=-{1\over 12\Phi^2}\, N^{\mu\nu\rho} N_{\mu\nu\rho}
-V\{\Phi\}\, ,\eqn{96b}\fe
while by (\ref{87}) and  (\ref{93}) the second part reduces to a 
2-dimensional string worldsheet integral given by
\be {\cal I}_{_{(2)}}=\int {\cal L}_{_{(2)}}\, d{\cal S}^{^{(2)}}
\, ,\hskip 1 cm {\cal L}_{_{(2)}}=-{\pi\,\hat l\over 2}\,\hbar^2\Phi^2 
- \pi\hbar B_{\mu\nu} {\cal E}^{\mu\nu}\, .\eqn{97b}\fe

The first term in (\ref{97b}) is like the corresponding action density
for a simple Nambu Goto (internally structureless) string, as
characterised by a string worldsheet stress-energy density tensor
$\overline T{^{\mu}_{\ \nu}}$ that is proportional to the fundamental
tangential projection tensor $\eta^\mu_{\ \nu}$ of the worldsheet as
defined by (\ref{90}), except that for an ordinary Nambu-Goto model the 
effective string tension ${\cal T}$ is fixed.  In the special limit case 
of the ``stiff'' fluid model (\ref{55}) which describes the massless axion 
case, it is well known~\cite{VilenkinVacha87} that the vortices will have 
an  effective tension ${\cal T}$ will also be fixed, but for a generic 
superfluid model one obtains 
\be \overline T{^{\mu}_{\ \nu}} =-{\cal T}\eta^\mu_{\ \nu}\, ,\hskip 1 cm
{\cal T}={\pi\hat l\over 2}\,\hbar^2\Phi^2\, ,\eqn{98}\fe
which shows that the effective tension ${\cal T}$ will vary
proportionally to the square of the auxiliary field amplitude, which according
to (\ref{62}) will be given on shell by 
\be \Phi^2={n\over\mu}\, .\eqn{100}\fe

The second term in (\ref{97b}) has the form of what is known in the context of
superstring theory as a ``Wess-Zumino'' coupling. When such a term turned up
in the ``stiff'' massless axion case (\ref{55}) it was at first
interpreted~\cite{VilenkinVacha87} as a ``very unusual interaction''.
However it soon came to be recognised ~\cite{DavisShellard89} as a
manifestation of an ordinary aerodynamic ``lift'' type force (arising from the
Magnus effect) of the kind first evaluated for an aerofoil in the long thin
(i.e. string type) limit by the Russian theorist Joukowski in the pionneering
days of subsonic flight a hundred years ago.

\section {The vortex string dynamical equations.}
\label{Section 10}

To work out the effects of the pair of terms in (\ref{97b}), one must obtain
the condition for the string type action ${\cal I}^{(2)}$ to be invariant with
respect to an infinitesimal displacement of the worldsheet generated by an
arbitrary vector field $k^\mu$ say. Any such displacement will give rise to
corresponding ``Lagrangian'' variations $\delta\Phi$, $\delta B_{\mu\nu}$,
$\delta g_{\mu\nu}$ (meaning variations as measured with respect to local
coordinates that are comoving with the displacement) of the relevant
background fields. Such ``Lagrangian'' variations will be given by
corresponding Lie differentiation formulae of the kind introduced above for
the respective cases of a scalar (\ref{21}), a closed antisymmetric covariant
field (\ref{27}), and last but in importance not least, the metric itself
(\ref{22}). Thus in the trivial case of the scalar $\Phi$ we shall simply have
\be \delta\Phi=k^\nu\nabla_{\!\nu}\Phi \, ,\eqn{101a}\fe
while in the case of the metric we have the familiar expression
\be \delta g_{\mu\nu}=2\nabla_{\!(\mu}k_{\nu)}\, ,\eqn{101b}\fe
(which would of course vanish if $k^\mu$ were not arbitrary but restricted to
be a Killing vector generating a background spacetime isometry, i.e. in
flatspace if $k^\mu$ were restricted to be a generator of some Poincar\'e
combination of translations and Lorentz transformations). Since the
antisymmetric field $B_{\mu\nu}$ is not closed (i.e. since its exterior
derivative, the current 3-form $N_{\mu\nu\rho}$ is non vavishing) the formula
for its Lie derivative is not quite so simple as that of its analogue
(\ref{27}) for the vorticity but contains an extra term: the relevant formula
is
\be \delta B_{\mu\nu}= k^\mu N_{\mu\nu\rho}
-2\nabla_{[\mu}(B_{\nu]\rho} k^\rho)\, .\eqn{101c}\fe
The corresponding variation of the string type action integral 
specified by (\ref{97b}) takes the standard form
\be\delta {\cal I}_{_{(2)}}=\int\big({_1\over^2}
\overline T{^{\nu\mu}}\delta g_{\mu\nu}+{_1\over^2}\overline W
{^{\nu\mu}}\delta B_{\mu\nu}+\overline{\cal F}\delta\Phi\big)\,
 d{\cal S}^{^{(2)}}\, , \eqn{102}\fe
in which it can be seen, taking account of the variation
of the surface element (\ref{85b}) due to the induced variation
\be \delta \Vert\gamma\Vert^{1/2}=
{_1\over^2}\Vert\gamma\Vert\eta^{\mu\nu}\delta g_{\mu\nu}\, ,\eqn{103}\fe
that the relevant surface stress energy momentum tensor
$\overline T{^{\mu\nu}}$ can be read out in the form (\ref{98}), and
that the relevant surface vorticity flux bi-vector will evidently
be as given by (\ref{93}) and (\ref{94b}), while finally the dynamical
dual of the amplitude $\Phi$ can be read out simply as
\be \overline{\cal F}= -\pi\,\hat l\, \hbar^2\,\Phi\, .\eqn{104}\fe

Since we are only concerned with local variations, there will be no
boundary contribution when we perform the usual operation of
integration by parts using Green's theorem so as to eliminate gradients of 
the arbitrary dispacement vector $k^\mu$. The result is thus obtained
in the standard form
\be\delta {\cal I}_{_{(2)}}=\int  k^\mu\big( \overline f_\mu
-\overline\nabla_{\!\nu}\overline T{^\nu}_{\!\mu}\big)\, d{\cal S}^{^{(2)}}
\eqn{105}\fe
in which, after taking account of the surface vorticity flux conservation 
law (\ref{91}), the effective surface force density acting in the 
ensuing dynamical equation,
\be \overline\nabla_{\!\nu}\overline T{^\nu}_{\!\mu}=\overline f_\mu\, ,
\eqn{106}\fe
can be read out as
\be \overline f_\mu={_1\over^2}N_{\mu\nu\rho}\overline W^{\nu\rho}
+ \overline {\cal F}\nabla_{\!\nu}\Phi\, .\eqn{107}\fe

In a model of this particular type (as a consequence of the fact that the
vortex string has no internal degrees of freedom of its own) the tangentially
projected part of the force balance equation (\ref{106}), i.e. the part that
is obtained by contracting it with the fundamental tensor $\eta^\mu_{\ \rho}$
given by (\ref{90}), will automatically be satisfied as a kinematic identity.
The only dynamical information in (\ref{106}) is the part obtained by
contracting it with the complement of the tangential projection tensor
$\eta^\mu_{\ \rho}$, namely the orthogonal projection tensor defined by
\be \perp^{\!\mu}_{\,\rho}= g^\mu_{\ \rho}-\eta^\mu_{\ \rho}\, .\eqn{108}
\fe

It can be seen that the remaining, purely orthogonal part of the left hand 
side of (\ref{106}) can be evaluated in terms of the extrinsic geometry of 
the imbedding, without knowledge of the gradient of $\overline T{^{\mu\nu}}$,
using the fact that, since the surface stress momentum energy tensor
$\overline T{^{\mu\nu}}$ is purely tangential, i.e. since its contraction with
$\perp^{\!\mu}_{\,\rho}$ vanishes, an integration by parts leads to the
identity
\be\perp^{\!\rho}_{\,\mu} \overline\nabla_{\!\nu}\overline T{^{\nu\mu}}
=\overline T{^{\mu\nu}} K_{\mu\nu}{^\rho}\, ,\eqn{109}\fe
where the  $K_{\mu\nu}{^\rho}$ is the second fundamental tensor
as defined~\cite{Carter89b} in terms of tangential differentiation of
the first fundamental tensor $\eta^\rho_{\ \sigma}$ by
\be K_{\mu\nu}{^\rho} = \eta^\sigma_{\ \nu}\overline\nabla_{\!\mu}
\eta^\rho_{\ \sigma} \, .\eqn{110}\fe
One thus obtains an expression of the standard form~\cite{Carter95}
\be \overline T{^{\mu\nu}} K_{\mu\nu}{^\rho}
=\perp^{\!\rho\mu}\overline f_\mu \, ,\eqn{111}\fe
for the extrinsic part of the dynamical equation (\ref{106}),
i.e. the part that governs the evolution of the worldsheet itself,
which is the only part there is in the present case.

Due to the Nambu-Goto like form (\ref{98}) obtained for $\overline
T{^{\mu\nu}}$ in the simple vortex model under consideration here, it will not
be necessary to work out the complete second fundamental tensor
$K_{\mu\nu}{^\rho}$ in the present case, but only its trace, the curvature
vector
\be K^\rho=K^\nu_{\ \nu}{^\rho}=\overline\nabla_{\!\nu}\eta^{\nu\rho}
\, ,\eqn{112}\fe
in terms of which it can be seen that the equation of motion (\ref{111})
will reduce to the form
\be {\cal T}K_\rho=\perp^{\!\nu}_{\,\rho}\nabla_{\!\nu}{\cal T}+
\pi\hbar{\cal E}^{\mu\nu}N_{\mu\nu\rho}\, ,\eqn{113}\fe
with the effective tension ${\cal T}$ given by (\ref{98}), according to which
it is proportional to the square of the auxiliary field $\Phi$ in the ambient
superfluid background. The first term on the right of (\ref{113}), allowing
for the effect of any non-uniformity of this field, was not needed in the
special limit of the ``stiff'' fluid model (\ref{55}) characterised by a fixed
value of $\Phi$, which describes the massless axion case to which earlier
studies of relativistic vortex string
dynamics~\cite{VilenkinVacha87,DavisShellard89} were restricted. 

\section{Vorton equilibrium in stationary background}
\label{Section 11}

As a simple exercise, let us apply the foregoing theory  to the problem of
a vorton, i.e. an equilibrium state of a vortex ring in a uniform background. 
As in the well known non relativistic version of this problem, relative motion 
is needed to provide the Joukowsky type force (due to the Magnus effect) that
supports the string against its own tension. A superfluid vorton differs from
local cosmic vorton~\cite{DavisShellard88,Carter95} in that the latter is 
supported not by a Magnus effect, but by the centrifugal effect of
circulating current.

By definition, an equilibrium state is invariant with respect to
the action of a time displacement vector, $k^\mu$ say, which, to preserve 
the background metric must satisfy the Killing equation given according to 
(\ref{22}) by
$2\nabla_{\!(\mu}k_{\nu)}=0$. This holds as a triviality  
for an ordinary flat space time displacement vector, which satisfies
\be\nabla_{\!\mu}k_{\nu}=0\, .\eqn{A1}\fe 
Invariance requires that the vortex 
worldsheet be tangent to this vector, whose components
in a corresponding Minkowski coordinate system will be given
by $k^\mu\leftrightarrow\{1,\, 0, \, 0, \, 0\}$ with normalisation
$k^\mu k_\mu=-c^2$.

The condition that the superfluid background is not just stationary but
uniform with velocity $v$ relative to such a  frame means that for suitably
aligned space axes its 4-velocity will be given by $u^\mu\ \leftrightarrow$
$\, \gamma\{1,\, v,\, 0,\, 0\}$ with $\gamma=(1-v^2/c^2)^{-1/2}$.
The assumption that the ambient fluid is uniform means that the gradient term
will drop out of the dynamical equation (\ref{113}), with the consequence that
the background number density $n$ will also cancel out, leaving an equilibrium
condition that is expressible just in terms of the 4-momentum covector
$\pi_\nu=\mu u_\nu$ in the form \be \hbar\, \hat l\, K_\rho=2{\cal
E}^{\mu\nu}\varepsilon_{\mu\nu\rho\sigma} \pi^\sigma\, .\eqn{A2}\fe

In such a stationary case it is easy to evaluate the tangent bivector ${\cal
E}^{\mu\nu}$ and the worldsheet curvature vector $K^\mu$ in terms of the unit
tangent vector, $e^\mu$ say, that is orthogonal to $k^\mu$ within the
worldsheet (which to be uniquely specified requires a choice of orientation).
In terms of this vector, as characterised by $e^\mu k_\mu=0$, $e^\mu e_\mu=1$,
we shall simply have
\be{\cal E}^{\mu\nu}= 2k^{[\mu} e^{\nu]}\, ,\eqn{A3}\fe
and (independently of the choice of orientation) the
corresponding  fundamental tensor  
will be given by 
\be \eta^{\mu\nu} =-{1\over c^2} u^\mu u^\nu +e^\mu e^\nu
\, .\eqn{A4}\fe
It can be seen from the defining relations (\ref{89}) and (\ref{112})
that when (\ref{A1}) is satisfied we shall simply have
\be K^\mu =e^\nu\nabla_{\!\nu} e^\mu\, .\eqn{A5}\fe

The vorton equilibrium states in which we are interested will be
symmetric about an axis aligned with the relative flow, which we take
to be the $x_{_1}$ axis of our Minkowski coordinate system.  The
worldsheet will be circular, with fixed radius $r$ say, meaning that
any point on it will be located relative to the axis by a radial
displacment vector that can be parametrised in terms of the
corresponding axial angle $\theta$ in the form $r^\mu\ \leftrightarrow$
$\ r\{0, \, 0,\ ,{\rm cos}\,\theta,\, {\rm sin}\,\theta\}$ with $r^\mu
r_\mu=r^2$. The corresponding unit tangent vector will be given by
$e^\mu\ \leftrightarrow$ $\ \{0, \, 0,\ ,-{\rm sin}\,\theta,\,
 {\rm cos}\,\theta\}$ and thus the corresponding curvature vector is
easily seen to be given by $K^\rho=r^{-2} r^\mu\ \leftrightarrow$
$\ r^{-1} \{0, \, 0,\ ,{\rm cos}\,\theta,\, {\rm sin}\,\theta\}$.

From this familiar result, namely that the curvature vector $K^\rho$ of
a circle of radius $r$ is inwardly directed with magnitude $r^{-1}$, it
can be seen that the solution to the equilibrium condition $(\ref{A2})$
will be obtained for a vorton radius given by
\be r ={\hbar\, \hat l\over 4\mu v\gamma}\, ,\eqn{A6}\fe
in which it is to be recalled that, according to (\ref{81}), $\hat l$
is an order of unity factor that can be estimated as the (natural)
logarithm of the ratio of the vorton area to the sectional area of the
vortex core.

The  vagueness of  the prescription for the $\hat l$ symptomises an
inherent limitation of any attempt (whether in a relativistic or
in a Newtonian framework) to treat a ``global'' vortex as if it were a
locally confined string type phenomenon, despite the ``infra-red
divergence'' of its energy and tension, which can only be made finite
by a long range cut off.  This problem does not arise for the vortex
defects in an ordinary metallic type II superconductor, in which the
local gauge coupling to the electromagnetic field ensures an
exponential fall off ensuring convergence without any need for a cut
off. In the cosmic string context, there are locally gauged examples
for which a string type description is highly accurate there are also
examples of non-local vortex defects, such as the axionic
case~\cite{VilenkinVacha87} (and cases involving
electromagnetic~\cite{Carter97} or gravitational~\cite{CarterBattye98}
coupling) for which the thin string limit description is less
satisfactory. If the vortex core radius $\delta$ is relatively small,
the problem of sensitivity to the long range cut off $\Delta$ is
mitigated by the fact that the dependence is only logarithmic. Thus a
formula such as (\ref{A6}) (or its well known non-relativistic
anologue, in which the $\gamma$ factor is omitted, and $\mu$ is
replaced by a fixed mass $m$) can be usefully applied to macroscopic
vorton configurations, but becomes quantitatively unreliable for
describing the microscopic vortons, known as ``rotons'', that are
important in the analysis of thermal excitations.

Whereas the problem of specification of the cut off can limit the utility of 
the string type description (\ref{97b}) for application to the dynamics of 
individual vortices, this reservation does not apply to the macroscopically 
averaged description for an Abrikosov type lattice of aligned vortices as 
given by  (\ref{83}), which is on a much sounder footing: in this case the 
relevant cut off is unambiguously determined by the lattice spacing. Although 
its accuracy is more questionable, the single string picture provides useful 
insight into the intepretation of the -- at first sight rather mysterious -- 
dynamical equation (\ref{82b}) for the more robust macroscopically averaged 
model, which is in fact formally identical to the more intuitively 
interpretable string dynamical equations (\ref{113}). 

\vfill\eject

\end{document}